\documentclass[aps,prc,reprint,superscriptaddress]{revtex4-2}
\usepackage{graphicx}
\usepackage{dcolumn}
\usepackage{bm}
\usepackage{amsmath}

\begin{document}


\title{Experimental study of $\alpha$-induced reactions on $^{113}$In for astrophysical $p$-process}


\author{Dipali Basak}
\email[]{dipali.basak@saha.ac.in}
\author{Tanmoy Bar}
\affiliation{Saha Institute of Nuclear Physics,1/AF, Bidhannagar, Kolkata$-$700064, India}
\affiliation{Homi Bhabha National Institute, Anushaktinagar, Mumbai$-$400094, India}

\author{Abhijit Roy}
\affiliation{Aryabhatta Research Institute of Observational Sciences, Manora peak, Uttarakhand$-$263001, India}

\author{Lalit Kumar Sahoo}

\author{Sukhendu Saha}
\affiliation{Saha Institute of Nuclear Physics,1/AF, Bidhannagar, Kolkata$-$700064, India}
\affiliation{Homi Bhabha National Institute, Anushaktinagar, Mumbai$-$400094, India}

\author{Jagannath Datta}
\author{Sandipan Dasgupta}
\affiliation{Analyical Chemistry Division, Bhabha Atomic Research Centre; Variable Energy
Cyclotron Centre, 1/AF Bidhannagar,
Kolkata$-$700064, India }

\author{Chinmay Basu}
\affiliation{Saha Institute of Nuclear Physics,1/AF, Bidhannagar, Kolkata$-$700064, India}
\affiliation{Homi Bhabha National Institute, Anushaktinagar, Mumbai$-$400094, India}
%

\date{\today}

\begin{abstract}
Neutron deficient nuclei from $^{74}$Se$-^{196}$Hg are thought to be produced by $\gamma$-induced reactions ($\gamma$,n), ($\gamma$,p) and ($\gamma,\alpha$) processes. The relatively high abundance of $^{113}$In odd A $p$-nuclei has inspired to study its production processes. As reaction with $\gamma$-beam is difficult to perform in the laboratory, $\gamma$-induced reaction rate is calculated from the inverse reaction data employing reciprocity theorem. Stacked foil activation method was used to measure the $^{113}$In($\alpha,\gamma$) and $^{113}$In($\alpha$, n) reactions cross-section near the astrophysical energies. Theoretical statistical model calculations were performed with different nuclear input parameters and compared with the experimental results. An appropriate $\alpha$-optical potential has been identified from the ($\alpha,\gamma$) and ($\alpha$, n) fitting, which provides the major source of uncertainty in the statistical model calculations. The other nuclear input parameters like level density, and $\gamma$-ray strength function were also constrained for theoretical calculations. $^{113}$In($\alpha,\gamma$)$^{117}$Sb and $^{117}$Sb($\alpha,\gamma$)$^{113}$In reaction rates were calculated using best-fitted input parameters.
\end{abstract}


\maketitle

\section{Introduction} 
Nuclei heavier than iron (Z$>$26) are formed by $s$-, $r$- and $p$-processes. The majority of the heavy nuclei are synthesized by either slow ($s$) neutron capture process or rapid ($r$) neutron capture process~\cite{RevModPhys.83.157, argast2004neutron}. However, 35 stable neutron-deficient isotopes ($^{74}$Se$-^{196}$Hg), known as $p$-nuclei, with isotopic abundances of less than 1$\%$ are bypassed by the neutron capture $s$- or $r$-process. These nuclei are produced from the $s$- or $r$-seed nuclei through a series of photodisintegration reactions, ($\gamma$, n), ($\gamma$, p) and ($\gamma, \alpha$). These reactions occur in high temperatures~(T$_9$~=~2$-$3) and high $\gamma$-flux environments, such as O/Ne burning layer in Type II supernovae~\cite{rayet1990p, arnould2003p, woosley1978p}. Neutron-rich seed nuclei are initially shifted towards the neutron-deficient side through a series of ($\gamma$, n) reactions. As the neutron threshold increases, ($\gamma, \alpha$) and ($\gamma$, p) photodisintegration reactions become dominant over ($\gamma$, n) reaction and produce stable lower Z elements. Thus ($\gamma$, p) and ($\gamma, \alpha$) are the key reactions that produce the $p$-isotopes. Therefore, these two reactions play an important role in determining the abundance of $p$-nuclei.

A huge reaction network calculation that involves more than 10000 nuclear reactions and about 1000 stable or unstable nuclei is used to model the $p$-process nucleosynthesis. The associate reaction rates determined from the reaction cross-section are important ingredients for the network calculations. Generally, the theoretical Hauser-Feshbach (HF) statistical model calculation predicts most of the photodisintegration reaction rates involving unstable nuclei. Therefore, it is crucial to experimentally measure the $\gamma$-induced reactions (($\gamma$, p) and ($\gamma, \alpha$)) on stable isotopes in order to verify the reliability of the theoretical predictions. Moreover, to standardise the parameters of the statistical model so that it can be used to predict the cross-section of reactions that cannot be measured experimentally.
 
Only a few direct $\gamma$-induced reactions have been studied due to the sparse availability of $\gamma$-beam facilities~\cite{PhysRevC.81.055806, PhysRevC.63.055802}. Additionally, the target nucleus is in the ground state in direct measurements at the laboratory. However, thermally populated excited states are also present in the stellar environment, contributing to the reaction rate. In an alternative way, the $\gamma$-induced reaction cross-section can be measured by performing an inverse capture reaction and calculating the reaction rate by the reciprocity theorem~\cite{arnould2003p, holmes1976tables}. In recent decades, several capture reactions near the astrophysically relevant energy region have been measured using activation technique~\cite{PhysRevC.58.524, PhysRevC.64.065803, PhysRevC.68.055803, ozkan2002cross, PhysRevC.66.015803, PhysRevC.74.025805, PhysRevC.75.025801, PhysRevC.79.065801, somorjai1998experimental, PhysRevC.78.035803}. The measured (p, $\gamma$) reaction cross-section generally agrees with the theoretical predictions within a factor of 2. However, there is a significant discrepancy between the measured value and the theoretical calculation for ($\alpha,\gamma$) reactions. 

\begin{figure}[h]
\includegraphics[scale=0.36]{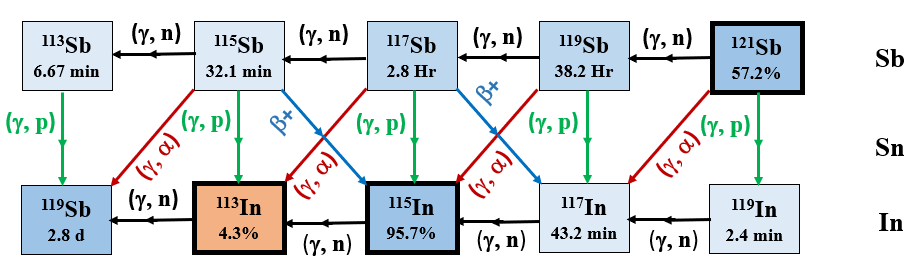}%
 \caption{\label{fig1}%
  The reaction pathway of the astrophysical $\gamma$-process for producing $^{113}$In $p$-nuclei. Only odd A isotopes are shown to simplify the reaction path. The double squares signify stable isotopes, while the single squares indicate unstable isotopes.
 }%
\end{figure}

The main objective of the current work is to measure the $^{113}$In($\alpha,\gamma$) reaction cross-section near the astrophysically important energy region. $^{113}$In is an odd-A $p$-nucleus (only two $p$-isotopes have odd mass numbers: $^{113}$In and $^{115}$Sn) and has a higher isotopic abundance (4.28$\%$) than many other $p$-isotopes. Furthermore, based on previous studies, it has been found that the $r$-process also contributes to the production of $^{113}$In~\cite{nemeth1994nucleosynthesis}. Therefore, an extensive investigation of production processes of $^{113}$In is important from an astrophysical perspective. A possible reaction path for the production of $^{113}$In $p$-nuclei is shown in Fig.~\ref{fig1}. The Gamow window for $^{113}$In($\alpha,\gamma$) reaction at T$_9$ = 3 is 6.13$-$9.03 MeV~\cite{PhysRevC.81.045807}. In this work, $^{113}$In($\alpha,\gamma$) reaction cross-sections were measured using activation technique in the energy range of 8.46 to 17.84 MeV. Since the statistical model calculation depends on the various nuclear input parameters (optical model parameters, level densities, $\gamma$-ray strength functions, etc.), the measured ($\alpha,\gamma$) cross-section was compared with the theoretical calculations by varying different input parameters. 

Moreover, $^{113}$In($\alpha$, n) reaction was also measured between 9.94 to 17.84 MeV. Though $^{113}$In($\alpha$, n) reaction has no astrophysical significance, the $\alpha$-optical potential, which is the major source of uncertainty in the calculations of reaction rates, can be derived from the ($\alpha$, n) reaction cross-section~\cite{basak2022determination, PhysRevC.105.014602}. As elastic scattering at astrophysical energies is dominated by Rutherford scattering, ($\alpha$, n) reaction is used for determining the  $\alpha$-optical potential.  Thus, a detailed study of $^{113}$In($\alpha,\gamma$) and $^{113}$In($\alpha$, n) reactions have been carried out to better understand the discrepancy between the experimental measurements and theoretical predictions. 
\section{Experimental procedure}
The product nuclei of $^{113}$In($\alpha,\gamma$)$^{117}$Sb and $^{113}$In($\alpha$, n)$^{116}$Sb reactions are both radioactive and decay by electron capture. Furthermore, the $^{113}$In($\alpha$, n) reaction populates the ground state of $^{116}$Sb (t$_{1/2}$ = 15.8 minute~\cite{firestone19978th, nndc}) as well as its isomeric state (t$_{1/2}$ = 60.3 minute~\cite{firestone19978th, nndc}). Only the partial reaction cross-section of the $^{113}$In($\alpha$, n)$^{116}$Sb$^{m}$ was measured due to the substantially longer half-life. The decay parameters of the product nucleus are listed in Table \ref{tbl1}. The reaction cross-sections were measured using the activation technique. The experiment was carried out at the K-130 Cyclotron facility, VECC, Kolkata. As the minimum available alpha beam energy from the K-130 Cyclotron facility was 28 MeV, the stacked foil technique was used in target irradiation setup. The targets after irradiation with the $\alpha$-beam were counted offline using HPGe $\gamma$-detectors. The different aspects of the experiment are discussed in the following subsections.

\begin{table*}[]
\caption{\label{tbl1} Decay parameters of the product nuclei from $\alpha$+$^{113}$In and $\alpha$+$^{nat}$Cu reaction~\cite{firestone19978th, nndc}.}
\begin{ruledtabular}
\begin{tabular}{cccc}
 Reaction and decay products & Half-life (minute) & $\gamma$-ray energy~~(keV) & $\gamma$-ray intensity I$_\gamma$ (\%) \\ 
 \hline
 \\
 \vspace{0.1cm}
 $^{113}$In($\alpha,\gamma$)$^{117}$Sb($\varepsilon$)$^{117}$Sn& 168.0$\pm$0.6 & 158.56 & 85.9\\
 $^{113}$In($\alpha$, n)$^{116}$Sb$^{m}$($\varepsilon$)$^{116}$Sn&60.3$\pm$0.6 & 407.35, 542.90 & 38.8, 52.0\\
 \hline
 $^{nat}$Cu($\alpha,x$)$^{67}$Ga$^{m}$($\varepsilon$)$^{67}$Zn&4697& 184.57 & 21.41\\
\end{tabular}
\end{ruledtabular}
\end{table*}

\subsection{Target preparation}
Targets made of both $^{nat}$In (4.29$\%$ $^{113}$In) and enriched $^{113}$In (93.7$\%$) were used in this experiment. The targets are prepared by vacuum evaporation on a 2.7 $\mu$m thick, 99.999$\%$ chemically pure Al foil. The In ingots were put on a Mo crucible, and the crucible was heated via bombardment with electron beams to evaporate the material. The backing Al foils were kept onto a substrate holder with a circular opening of 12 mm in diameter, and the holder was placed 16 cm above the Mo crucible. The thickness of the targets was obtained by measuring the energy loss of $\alpha$-particles from a known 3-line $\alpha$-source ($^{239}$Pu, $^{241}$Am, $^{244}$Cm ). To check the uniformity of the target, the thickness was measured at multiple positions on the target. The thickness of the prepared targets was between 60$-$85 $\mu$g/cm$^2$. The uncertainty of the target thickness is about 6$-$7 $\%$. 

In the activation measurement, the final reaction residue has to be stopped in the backing material. The reaction residue $^{117}$Sb has recoil energy around 620 keV at the highest $\alpha$-beam energy 18 MeV and stopped at 0.2~$\mu$m Al foil, which is much lower than the backing Al thickness (2.7 $\mu$m). Maximum recoil energy of $^{116}$Sb was 1.13 MeV at maximum $\alpha$-beam energy and 0.5~$\mu$m Al foil is sufficient to stop the $^{116}$Sb residue.

\subsection{Stacked target irradiation}
A stack made up of several targets was irradiated in order to measure the reaction cross-section at multiple energy points simultaneously~\cite{chowdhury1995determination, takacs2013cross, PhysRevC.92.064611}. In the first type, the stack consisted of two In targets and a 7$\mu$m Cu foil placed in between the targets~(type 1). The stack is formed with one In target and a Cu foil in the other type~(type 2). Cu foil is used as a monitor foil to measure the beam current from the known $^{nat}$Cu($\alpha,x$)$^{67}$Ga reaction cross-section~\cite{shahid2015measurement, takacs2017crosschecking, tarkanyi2000new}. The Al foil (6 $\mu$m) after the Cu is used as a catcher foil for Cu target and 1 mm thick Al ring was used as a spacer between two foils. Out of seven target stacks, three stacks contained two enriched $^{113}$In targets, while the remaining four stacks contained one $^{113}$In target, were irradiated with $\alpha$-beam. $^{nat}$In targets were also irradiated using three target stacks. One stack consisted of two $^{nat}$In targets, whereas the other two contained a single $^{nat}$In target. Each target stack was irradiated for 2.5 to 18 hour, depending on the longest half-life of the reaction products and the beam energy. The type 1 stacks were irradiated at higher energies, while type 2 stacks were irradiated at lower energies. The type 2 stacks were irradiated for a longer period of time to get enough statistics since the reaction cross-section is abruptly decreased at lower energies.

\begin{figure*}
\includegraphics[scale=0.9]{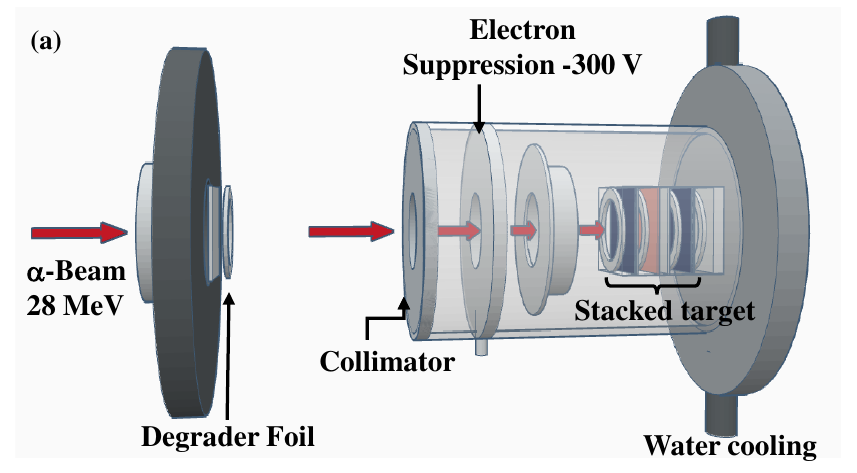}%
\quad
\includegraphics[scale=0.72]{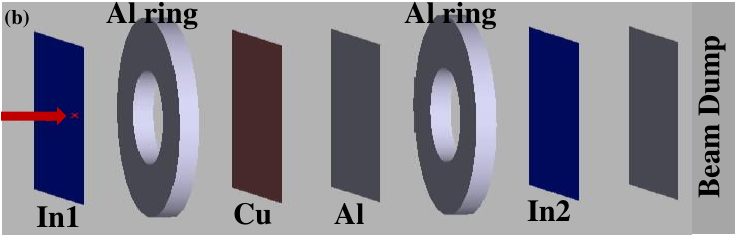}%
 \qquad
\includegraphics[scale=0.89]{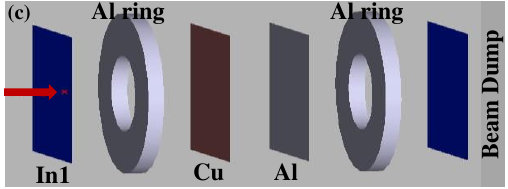}
 \caption{\label{fig2}%
  Schematic diagram of the experimental setup used in stacked foil irradiation 
   (a): Irradiation chamber 
   (b): Stack contains two In targets (Type 1)
   (c): Stack contains one In target (Type 2).
 }%
\end{figure*}

The average current of He$^{2+}$ beam was around $\sim$250 nA. A current integrator was connected to the irradiation chamber, which recorded the current in 0.3 second intervals. A suppression voltage ($-$300 V) was applied to prevent the secondary electrons from escaping the Faraday cup, which can otherwise register a higher beam current. It is also possible that some portion of the beam may strike the chamber wall instead of hitting the targets entirely. In both the cases recorded current may be misleading. Monitor (Cu) foil is used to validate the beam current. Al degrader foils were placed just before the irradiation chamber to reduce the $\alpha$-beam energy. The thickness of the Al degrader foils ranged from 184 to 296 $\mu$m in order to measure the cross-sections at various energies. Fig.~\ref{fig2} depicts the schematic design of the irradiation chamber and target stacks.

\begin{figure}[h]
\includegraphics[scale=0.42]{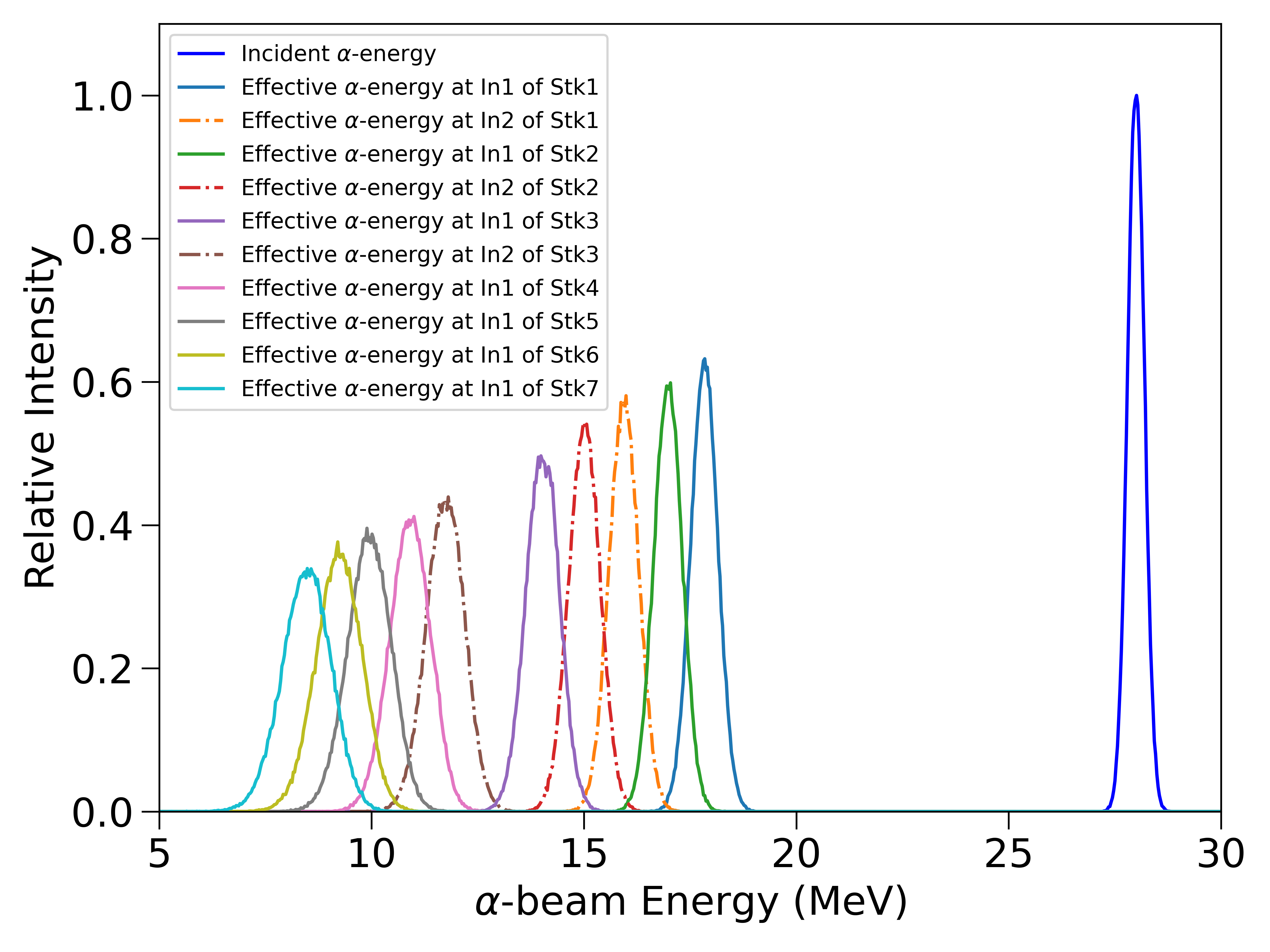}%
 \caption{\label{fig3}%
 Effective $\alpha$-beam energy with energy straggling for enrich $^{113}$In targets simulated from GEANT4. Stk1$-$Stk3 are consisting of two $^{113}$In targets and Stk4$-$Stk7 consisting of one $^{113}$In target. 
 }%
\end{figure}

The crucial part of this measurement is to determine the $\alpha$-beam energy in each target of each stack. The energy loss and straggling of $\alpha$-beam in each target have been calculated via the Monte Carlo simulation GEANT4 toolkit~\cite{agostinelli2003geant4}. The initial $\alpha$-beam energy from the Cyclotron is 28 MeV with $\sim$200 keV energy resolution. The total uncertainty in the interaction energy of the $\alpha$-beam is caused by the initial uncertainty in the beam energy, the energy spread, and the uncertainty in the foil thickness. The interaction energies with energy straggling in each $^{113}$In target are shown in Fig.~\ref{fig3} .  

\subsection{$\gamma$-ray detection}
\begin{figure}[hb]
\includegraphics[scale=0.34]{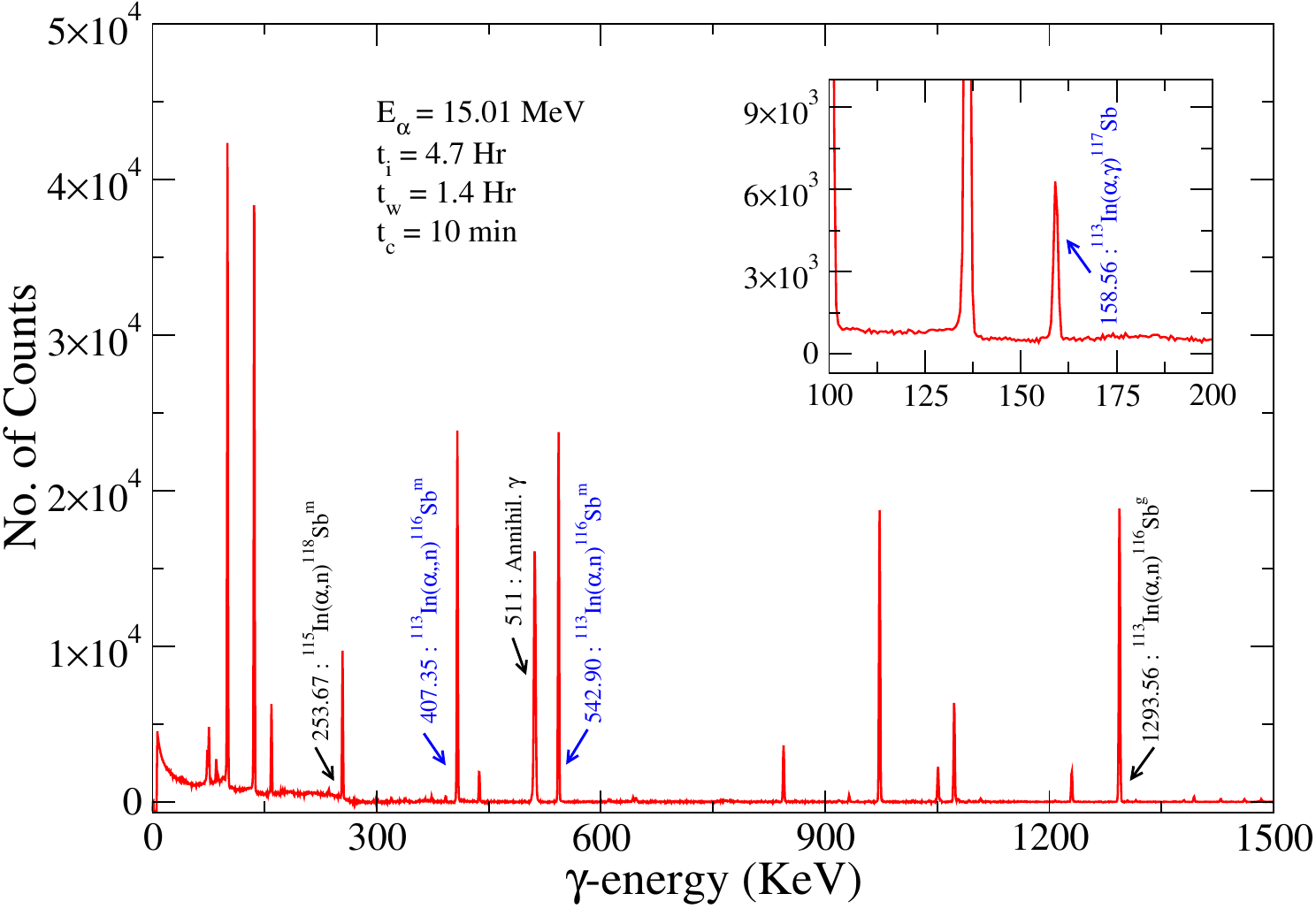}%
 \caption{\label{fig4}%
  $\gamma$-ray spectrum for the enriched $^{113}$In target irradiated with 15.01 MeV $\alpha$-beam. The necessary $\gamma$-lines for cross-section measurements are marked by the blue arrows. Other $\gamma$-lines are coming from $^{115}$In impurity in the target material and laboratory background. 
 }%
\end{figure}

The target stack was brought out from the chamber and kept to cool for about an hour after every irradiation. After cooling, the activity produced by each target is measured using an HPGe detector having a relative efficiency of 40$\%$.  Lead blocks (7.5 cm thick) have been used to shield the detector in order to decrease the background. Indium targets that had been irradiated with higher energy alpha beams were put at a 37.5 mm distance from the HPGe detector end cap for $\gamma$-counting, whereas targets that had been irradiated to lower energy alpha were placed 12.5 mm away (close geometry). The $\gamma$-activity from monitor copper foils was detected at a distance of 37.5 mm from the detector window. The $^{113}$In($\alpha,\gamma$)$^{117}$Sb reaction cross-sections were measured by determining the activity of $^{117}$Sb (t$_{1/2}$ = 2.8 hour), that was derived from 158.56 keV $\gamma$-lines. Also, the $^{113}$In($\alpha$, n)$^{116}$Sb$^m$ reaction cross-sections have been determined from the $^{116}$Sb$^m$ (1.01 hour) activity, which was obtained from the peak area at 407.35 or 542.9 keV in the $\gamma$-spectrum. $\gamma$-spectra were counted for 10 minutes to 10 hours in order to accumulate enough statistics for the desired $\gamma$-peak. Fig.~\ref{fig4} shows the offline $\gamma$-spectrum for the In target irradiated with 15.01 MeV $\alpha$-beam for 4.7 hours and counted for 10 minutes after 1.4 hours cooling. Arrows correspond to the necessary $\gamma$-lines for cross-section measurements.

The efficiency and energy calibration of the HPGe detector were measured using a calibrated multi-energy source $^{152}$Eu at the 162.5 mm distance (far geometry) where coincidence summing effect is negligible. The coincident summing effect needs to be taken into consideration for close geometry measurements~(12.5 mm and 37.5 mm). One extra $^{nat}$In target was irradiated at 17.41 MeV, and the required $\gamma$-lines yields were measured in both
close and far geometries. The efficiency conversion factor was obtained by taking the ratio of the $\gamma$-yield in close geometry to that in far geometry. Thus, the efficiency of the detector was modified by multiplying the factor to the detection efficiency at far geometry and used this in the close geometry calculations.

\section{Data Analysis}
\begin{figure*}[htp]
\includegraphics[scale=0.48]{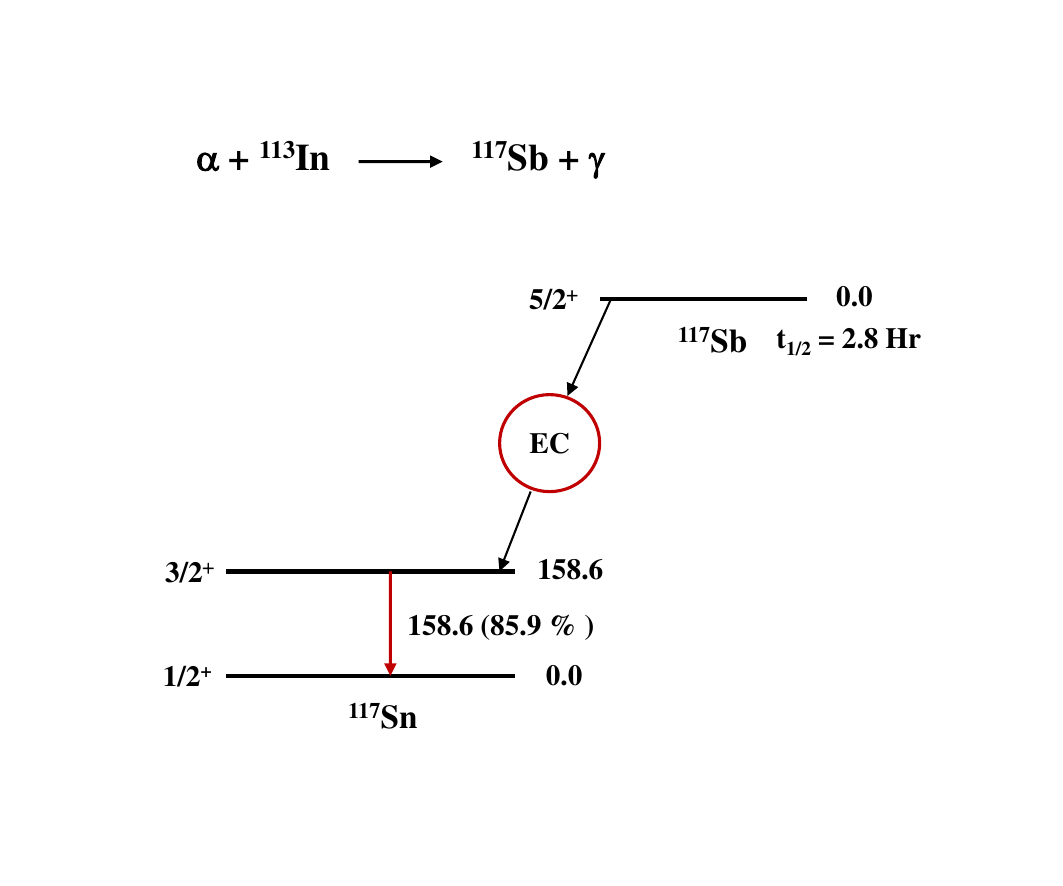}%
\quad
\includegraphics[scale=0.48]{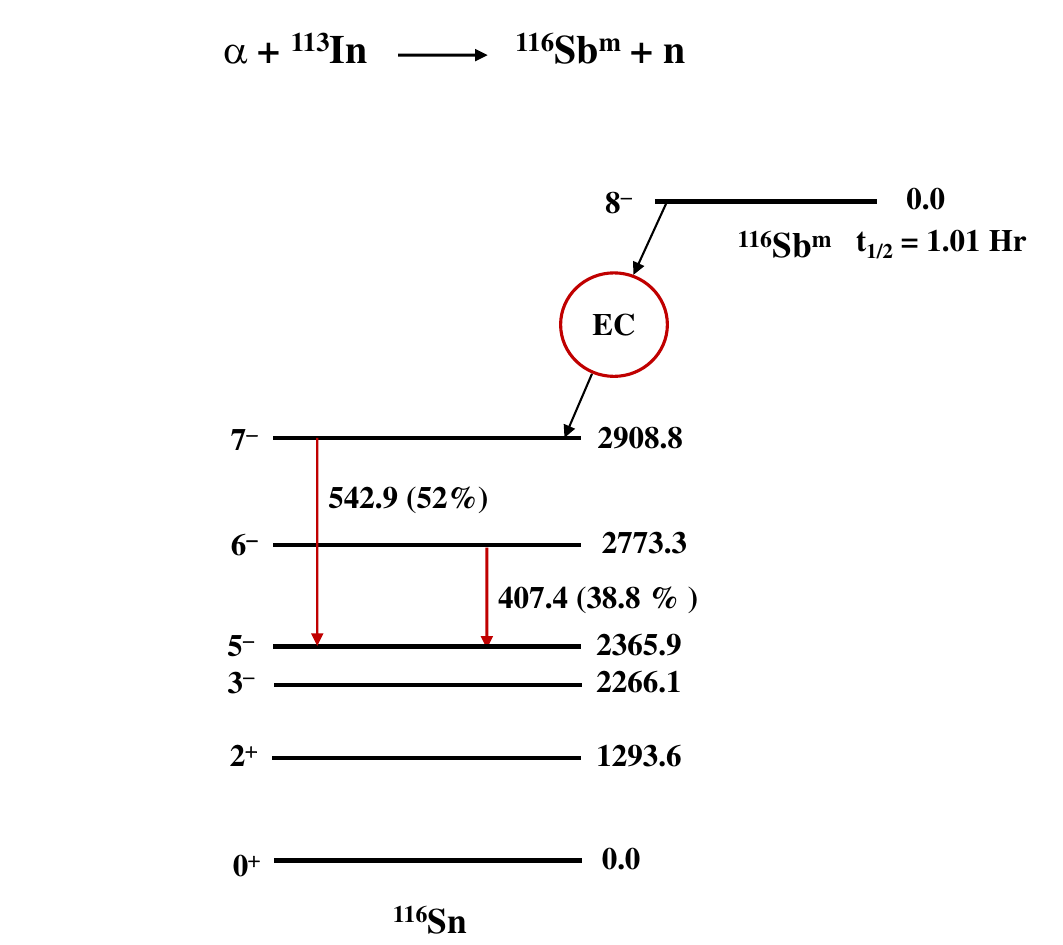}%
 \caption{\label{fig5}%
 Decay schemes of the $^{117}$Sb and $^{116}$Sb$^{m}$. Information of spin-parities, energies and $\gamma$-ray intensities are taken from~\cite{nndc}.
 }%
\end{figure*}

Through electron capture, the product nuclei $^{117}$Sb and $^{116}$Sb$^m$ from the $^{113}$In($\alpha,\gamma$) and $^{113}$In($\alpha$, n) reactions decay to $^{117}$Sn and $^{116}$Sn, respectively, with half-lives of t$_{1/2}$ = 2.8 hour and 1.01 hour. The decay products~($^{117}$Sn and $^{116}$Sn), that were in the excited state, de-excited to the ground state by the emission of $\gamma$-rays. The stronger $\gamma$-transitions were used for the cross-section measurements. The decay schemes of the reaction products $^{117}$Sb and $^{116}$Sb$^{m}$ are shown in Fig.~\ref{fig5}.

\begin{figure}[hb]
\includegraphics[scale=0.18]{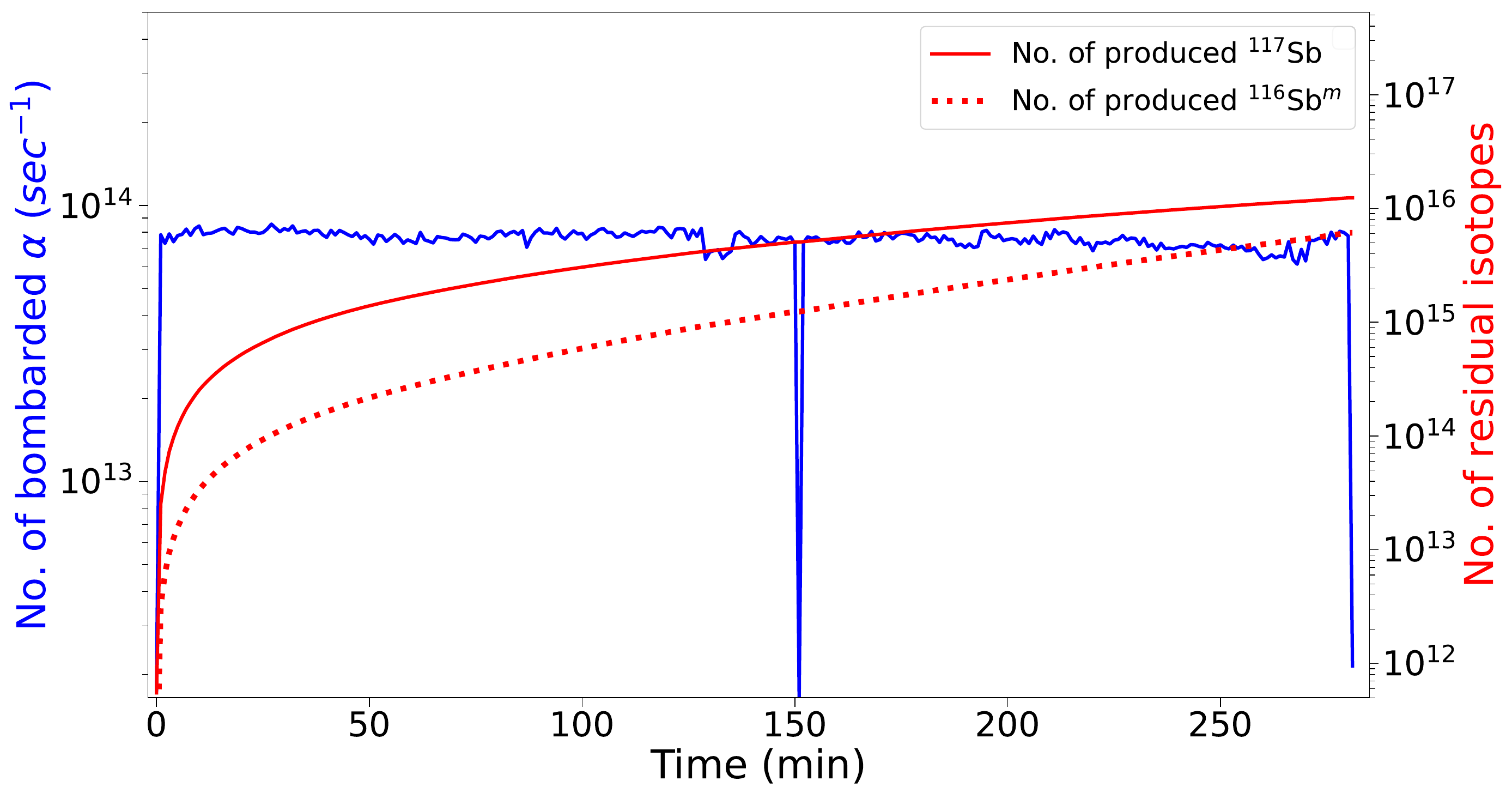}%
 \caption{\label{fig6}%
  Beam current vs number of residual nuclei produced from $^{113}$In($\alpha,\gamma$) and the $^{113}$In($\alpha$, n) reactions.
 }%
\end{figure}

The total number of decays of the residual nuclei in the counting interval can be determined by calculating the peak count (A$_{\gamma}$) in the $\gamma$-spectrum of a certain transition according to the given relation~\cite{kruger1971principles},

\begin{equation}\label{eq812}
N_{\rm decay} = \frac{\rm A_{\gamma}}{\epsilon I_{\gamma}}
\end{equation}
where $\epsilon$ is the efficiency of the detector and $I_{\gamma}$ be the intensity of the $\gamma$-transition. The peak count of the desired $\gamma$-lines was determined by deducting the background counts from the neighbouring higher and lower energy sides using the CERN ROOT data analysis tool~\cite{brun1997root}. The reaction cross-section was determined using, 

\begin{equation}
\sigma_{\rm reac} = \frac{\lambda \rm A_{\gamma}}{\epsilon I_{\gamma} {\rm N_A}\phi_b\left(1~-~e^{-\lambda t_i}\right)e^{-\lambda t_w}\left(1~-~e^{-\lambda t_c}\right)}
\end{equation}
where $\rm N_A$ is the surface density of the target (atoms/cm$^2$) and $\lambda$ be the decay constant that is related to the half-life ($t_{1/2}$) of residual nuclei as $t_{1/2}$ = ln(2)/$\lambda$. $t_w$ and $t_c$ are the cooling time and counting time, respectively. Since the beam current fluctuates throughout the irradiation, the total irradiation period ($t_i$) is divided into N intervals ($\Delta t$ = 1 minute), during which the beam current is assumed to be constant. Therefore, the cross-section formula will modify as~\cite{gyurky2019activation},

\begin{equation}
\sigma_{\rm reac} = \frac{\rm A_{\gamma}}{\epsilon I_{\gamma} {\rm N_A} e^{-\lambda t_w}\left(1~-~e^{-\lambda t_c}\right)\sum_{i=1}^{\rm N} \phi_{b,i}\frac{1-e^{-\lambda\Delta t}}{\lambda}e^{-\lambda\Delta t \left({\rm N}-i\right)}}
\end{equation}
where $\phi_{b,i}$ is the number of bombarded $\alpha$-particle per sec in the $i^{th}$ time period. The beam current profile during the irradiation for Stk2 of $^{113}$In is shown in Fig.~\ref{fig6}.

The presence of $^{115}$In in the enriched $^{113}$In and $^{nat}$In targets is 6.9$\%$ and 95.7$\%$, respectively. Therefore, both the $^{113}$In($\alpha,\gamma$) and the $^{115}$In ($\alpha$, 2n) reaction channels yield $^{117}$Sb radionuclide. Since the $^{115}$In ($\alpha$, 2n) reaction has a threshold energy of 15.01 MeV, above this energy, the $^{115}$In ($\alpha$, 2n) reaction channel also contributes to the production of $^{117}$Sb.  Therefore, the activity ratio measurement has to be used to determine the contribution of $^{117}$Sb activity from a particular reaction~\cite{SMITH198237, vashi2022cross}. The following relation was used for computing the activity ratio (R) of $^{117}$Sb from $^{113}$In($\alpha,\gamma$) and $^{115}$In ($\alpha$, 2n) reaction.

\begin{equation}
\begin{split}
\rm {R} & = \frac{\rm{Activity~of~^{117}Sb~from~^{113}In (\alpha, \gamma)}}{\rm{Activity~of~^{117}Sb~from~^{115}In (\alpha, 2n)}} 
\\
& = \frac{\rm{\sigma_{^{113}In(\alpha, \gamma)}N_{^{113}In}}}{\rm{\sigma_{^{115}In(\alpha, 2n)}N_{^{115}In}}}
\end{split}
\end{equation}
where N$_{\rm{^{113}In}}$ and N$_{\rm{^{115}In}}$ are the surface densities of the $^{113}$In and $^{115}$In isotopes in the target material, respectively. The reaction cross-sections $\sigma_{\rm ^{113}In(\alpha, \gamma)}$ and $\sigma_{\rm ^{115}In(\alpha, 2n)}$ are estimated using code TALYS-1.96~\cite{koning2023talys} with the default parameters. The activity of $^{117}$Sb from the specific reaction is calculated using the activity factor, and the overall activity is determined by the $\gamma$-lines in the $\gamma$-spectrum.

The quadratic sum of all systematic errors and statistical error was considered in determining the uncertainty in the reaction cross-section values. The source of statistical error is the peak count rate of the required $\gamma$-lines, that varies between 0.4$-$10 $\%$. Uncertainty in the measurements of target thickness (6$-$7$\%$), detector efficiency ($\sim$9$\%$), and decay parameters from literature ($\sim$5$\%$) are the primary source of systematic uncertainty in the cross-section. Uncertainty in the primary beam energy, uncertainty in the foil thickness and the beam straggling when passing through the target and degrader foil are the source of uncertainty in the effective beam energy.

\section{Results and Model predictions}
\subsection{Reaction cross-section}
$^{113}$In($\alpha,\gamma$) reaction cross-section was measured in the laboratory energy range 8.46$-$17.84 MeV from the 158.56 keV $\gamma$-line. At the lowest two energies~(8.46 and 9.22 MeV), the yield of the $^{113}$In($\alpha$, n) reaction channel becomes extremely poor. Therefore, ($\alpha$, n) reaction cross-section is measured in the energy range 9.94$-$17.84 MeV. The partial reaction cross-section of the $^{113}$In($\alpha$, n)$^{116}$Sb$^{m}$ was measured from the 407.35 and 542.9 keV $\gamma$-line. The measured cross-sections of $^{113}$In($\alpha,\gamma$) and $^{113}$In($\alpha$, n)$^{116}$Sb$^{m}$ reaction are listed in Table~\ref{tbl2} and Table~\ref{tbl3}, respectively. Experimental results obtained from enriched and natural targets have been reported separately. The cross-section values obtained in the current study are relatively lower than those obtained in previous measurements~\cite{PhysRevC.79.065801}.

\begin{table}[h!]
\caption{\label{tbl2}%
Experimental $^{113}$In($\alpha,\gamma$)$^{117}$Sb reaction cross-section
}
\begin{ruledtabular}
\begin{center}
\begin{tabular}{cc}
\textrm{Effective beam energy (MeV)}&
\textrm{cross-section (mbarn)}\\
\colrule
 \multicolumn{2}{c}{$^{113}$In target}     \\
17.84$\pm$0.33 & 0.9311$\pm$0.1138\\
16.98$\pm$0.35 & 1.0249$\pm$0.1253\\
15.94$\pm$0.36 & 0.7651$\pm$0.0936\\
15.01$\pm$0.38 & 0.7557$\pm$0.0925\\
14.02$\pm$0.41 & 0.7002$\pm$0.0857\\
11.73$\pm$0.45 & 0.0581$\pm$0.0072\\
10.90$\pm$0.48 & 0.0361$\pm$0.0067\\
9.94$\pm$0.52 & 0.0118$\pm$0.0017\\
9.22$\pm$0.55 & 0.0073$\pm$0.0009\\
8.46$\pm$0.58 & 0.0015$\pm$0.0003\\
 \multicolumn{2}{c}{$^{nat}$In target}     \\
17.41$\pm$0.34 & 0.6801$\pm$0.0829\\
16.53$\pm$0.36 & 1.0284$\pm$0.1255\\
14.44$\pm$0.39 & 0.6504$\pm$0.0863\\
13.02$\pm$0.42 & 0.2729$\pm$0.0367\\
\end{tabular}
\end{center}
\end{ruledtabular}
\end{table}

\begin{table}[b]
\caption{\label{tbl3}%
Experimental $^{113}$In($\alpha$, n)$^{117}$Sb reaction cross-section
}
\begin{ruledtabular}
\begin{center}
\begin{tabular}{cc}
\textrm{Effective beam energy (MeV)}&
\textrm{cross-section (mbarn)}\\
\colrule
 \multicolumn{2}{c}{$^{113}$In target}     \\
17.84$\pm$0.33 & 146.197$\pm$17.848\\
16.98$\pm$0.35 & 106.016$\pm$12.939\\
15.94$\pm$0.36 & 35.997$\pm$4.394\\
15.01$\pm$0.38 & 24.898$\pm$3.039\\
14.02$\pm$0.41 & 12.215$\pm$1.492\\
10.90$\pm$0.48 & 0.041$\pm$0.007\\
9.94$\pm$0.52 & 0.009$\pm$0.002\\
 \multicolumn{2}{c}{$^{nat}$In target}     \\
17.41$\pm$0.34 & 87.775$\pm$11.107\\
16.53$\pm$0.36 & 63.199$\pm$8.151\\
14.44$\pm$0.39 & 15.705$\pm$2.033\\
13.02$\pm$0.42 & 2.009$\pm$0.274\\
\end{tabular}
\end{center}
\end{ruledtabular}
\end{table}

\begin{figure}[h!]
\includegraphics[scale=0.34]{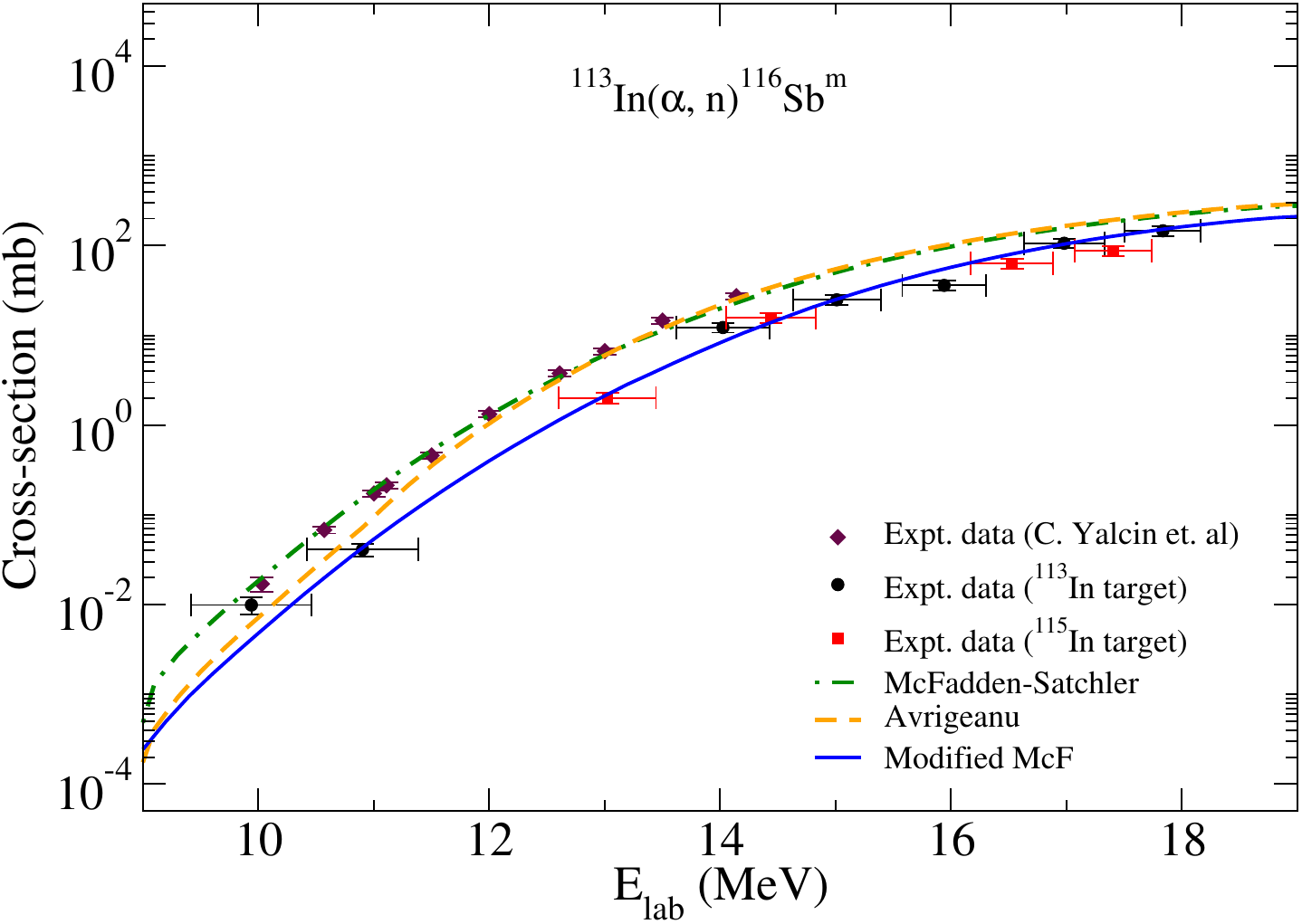}%
 \caption{\label{fig7}%
Measured $^{113}$In($\alpha$, n) reaction cross-section. Theoretical Hauser-Feshbach calculations have been performed using different $\alpha$-OMP.
 }%
\end{figure}

\begin{figure}[h!]
\includegraphics[scale=0.34]{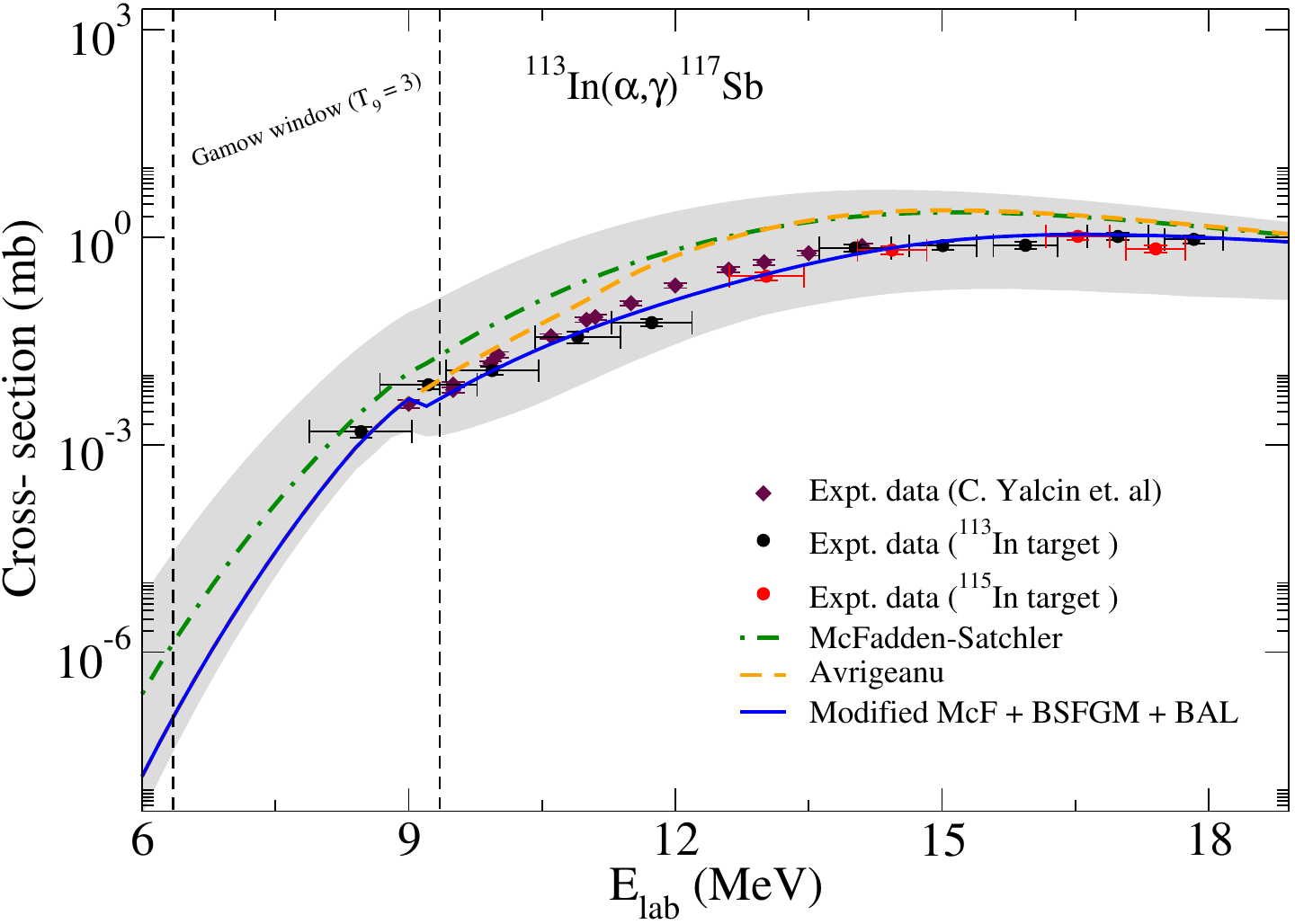}%
 \caption{\label{fig8}%
 Measured $^{113}$In($\alpha$, n) reaction cross-section. Theoretical Hauser-Feshbach calculations have been performed using different nuclear input parameters. The gray shaded region represents the range of probable cross-section by varying the nuclear input parameters ($\alpha$-OMP, LD, $\gamma$-SF) in TALYS code.
 }%
\end{figure}

Theoretical calculations have been performed using the Hauser-Feshbach statistical model code TALYS-1.96~\cite{koning2023talys} and compared with the measured values, as shown in Fig.~\ref{fig7} and Fig.~\ref{fig8}. The model calculations have been carried out with different combinations of nuclear inputs used in statistical model calculations ($\alpha$+nucleus optical potentials, level densities, $\gamma$-ray strength functions). Eight types of $\alpha$+nucleus optical model potentials, six types of level density models and nine types of $\gamma$-ray strength functions are included in the TALYS-1.96 code. It was found that the theoretical cross-sections calculated using the two well-known global $\alpha$+nucleus potentials McFadden-Satchler~\cite{mcfadden1966optical} and Avrigeanu~\cite{avrigeanu2014further} accurately predicted the previously measured ($\alpha$, n) reaction data but significantly overpredicted the present data. The calculations using modified McF potential from Ref.~\cite{basak2022determination} are in better agreement with the present measurements. The theoretical ($\alpha$, n) cross-sections are not significantly changed by varying the different level density model and $\gamma$-ray functions. However, modified McF potential adequately explain the ($\alpha,\gamma$) reaction data with Backshifted Fermi Gas model(BFM)~\cite{dilg1973level} for level density and Brink Axel Lorentzian(BAL)~\cite{axel1962electric, brink1957individual} for $\gamma$-ray strength function. In Hauser-Feshbach model calculation, reaction cross-sections are computed from the transmission coefficients that are correlated with the average width parameters. Transmission coefficients are obtained by solving the Schrodinger equation using interaction potentials for particle channels and $\gamma$-ray strength functions and level densities for photon channels.

\begin{figure}[h!]
\includegraphics[scale=0.35]{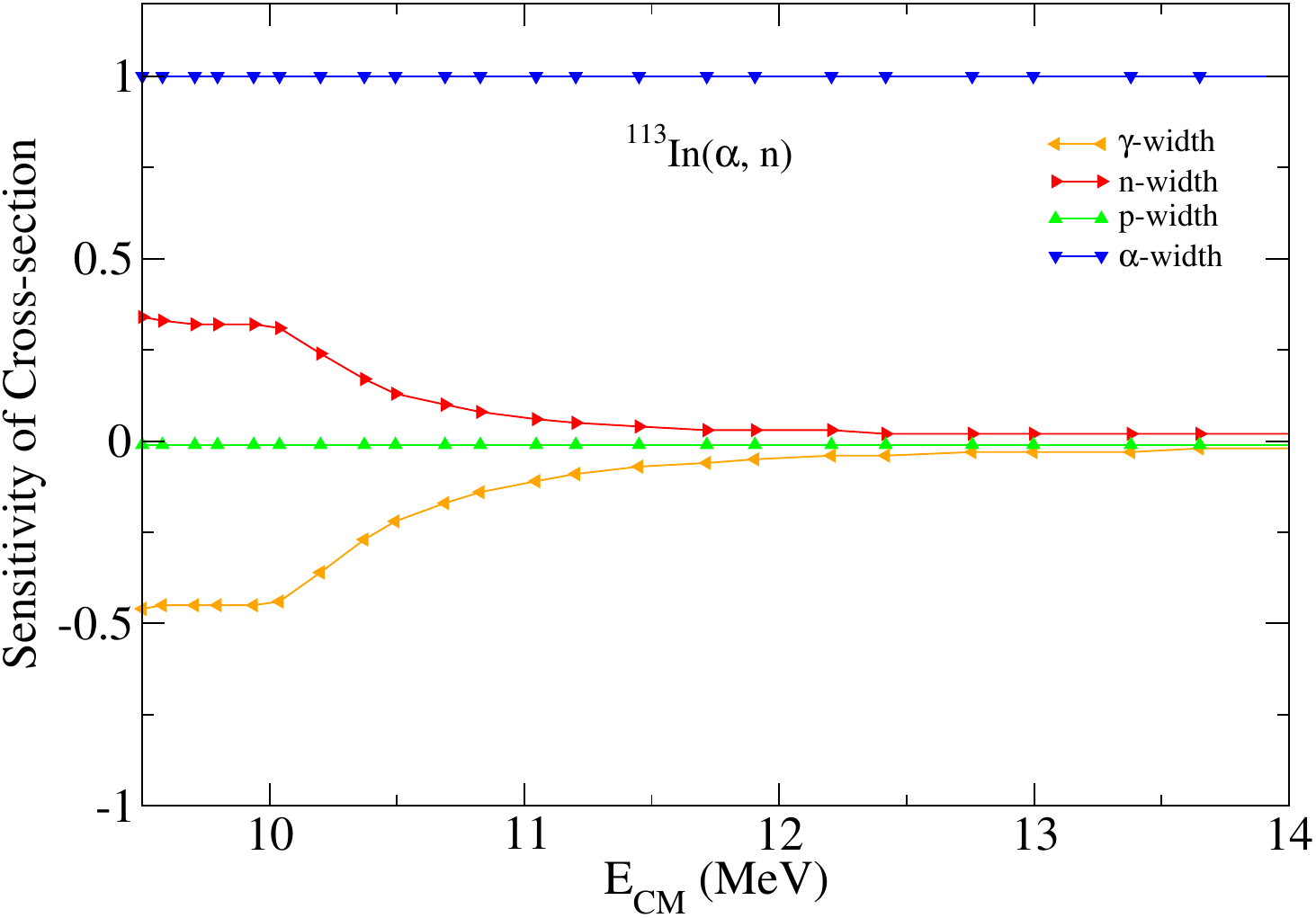}%
 \caption{\label{fig9}%
  $^{113}$In($\alpha$, n) reaction cross-section sensitivity with respect to center of mass energy by varying of $\alpha$-,n-,p- and $\gamma$-width. Sensitivity data are taken from~\cite{rauscher2012formalism}. 
 }%
\end{figure}

\begin{figure}[h!]
\includegraphics[scale=0.35]{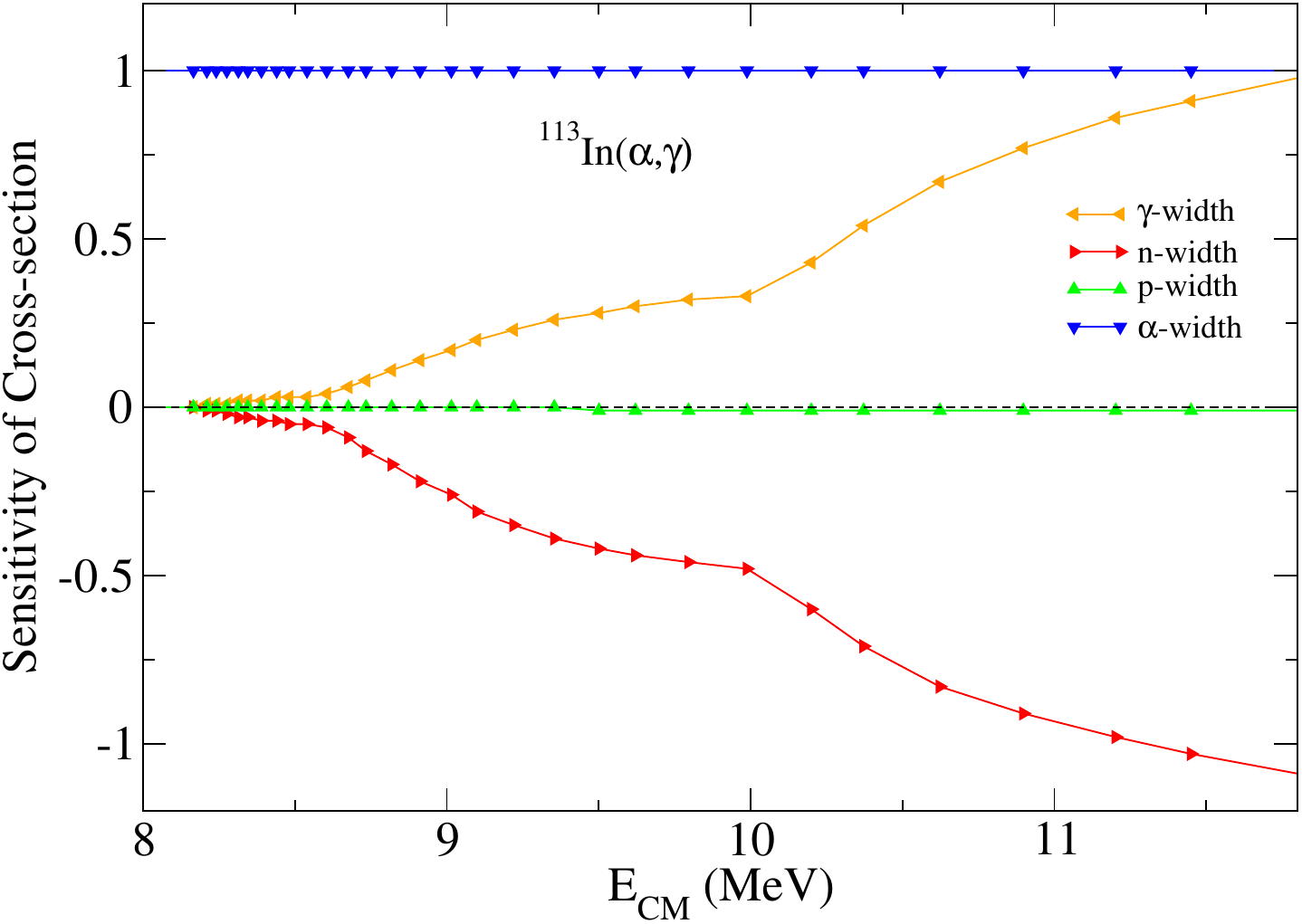}%
 \caption{\label{fig10}%
  Same as Fig.~\ref{fig9} except for $^{113}$In($\alpha,\gamma$) reaction.
 }%
\end{figure}

Therefore, it is important to take into account the sensitivity of the predicted cross-section on the nuclear input parameters used in the model computation in order to properly comprehend the discrepancy between theory and experiment. Thomas Rausher defined the sensitivity factor as~\cite{rauscher2012formalism},
\begin{equation}
S_{\sigma} = \frac{\frac{\rm d\sigma}{\sigma}}{\frac{{\rm d}w}{w}}
\end{equation}
where d$\sigma$ and dw are the changes in cross-section and width (particle or $\gamma$), respectively. $S_{\sigma}$ = $\pm$1 indicates a change in cross-section proportional to the change in width, whereas $S_{\sigma}$ = 0 implies no change in cross-section due to change in width. A complete analysis of reaction cross-section sensitivities is presented in Ref.~\cite{rauscher2012formalism}. The cross-section sensitivity of ($\alpha,\gamma$) and ($\alpha$, n) reactions with the change of various width parameters is shown in Fig.~\ref{fig9} and Fig.~\ref{fig10} respectively. The sensitivity plot showed that the impact of proton width is negligible for both reactions. 

In the measured energy region ($\alpha$, n) reaction cross-section is mainly sensitive to the $\alpha$-width parameter, i.e., $\alpha$-transmission coefficient. Thus $\alpha$+nucleus potential is only affected the ($\alpha$, n) reaction cross-section data in the model calculation. But the sensitivity of ($\alpha,\gamma$) reaction reaction is a little more complicated. ($\alpha,\gamma$) reaction cross-section is only sensitive to the $\alpha$-width parameter in the lower energy region (Gamow window),  but above the Gamow window, the $\alpha$-, n-, and $\gamma$-widths are all important factors for cross-section calculation. 

The ($\alpha$, n) reaction has less significance in astrophysics studies. The primary objective of this study is to evaluate the $\alpha$-optical potential. The measured ($\alpha$, n) reaction cross-section was suitably reproduced by the modified McF potential. Therefore, it can be concluded that this potential more precisely defines the $\alpha$-width for this system. The modified McF potential is basically the well-known energy and mass independent McF potential with both of its real and imaginary width parameters modified by an energy dependent Fermi function. The ($\alpha,\gamma$) reaction cross-section is computed using the modified McF potential with different level density and $\gamma$-ray strength function combinations. The ($\alpha,\gamma$) reaction cross-section are constantly affected by the $\alpha$-width parameter in the measured energy range. However, the sensitivities of n- and $\gamma$-width increases with increasing energy but in opposite manner. The modified McF potential reproduces the ($\alpha,\gamma$) reaction cross-section data with BFM for level density and BAL for $\gamma$-ray strength function. 

\subsection{Astrophysical $S$-factor and stellar reaction rate}
\begin{figure}[]
\includegraphics[scale=0.35]{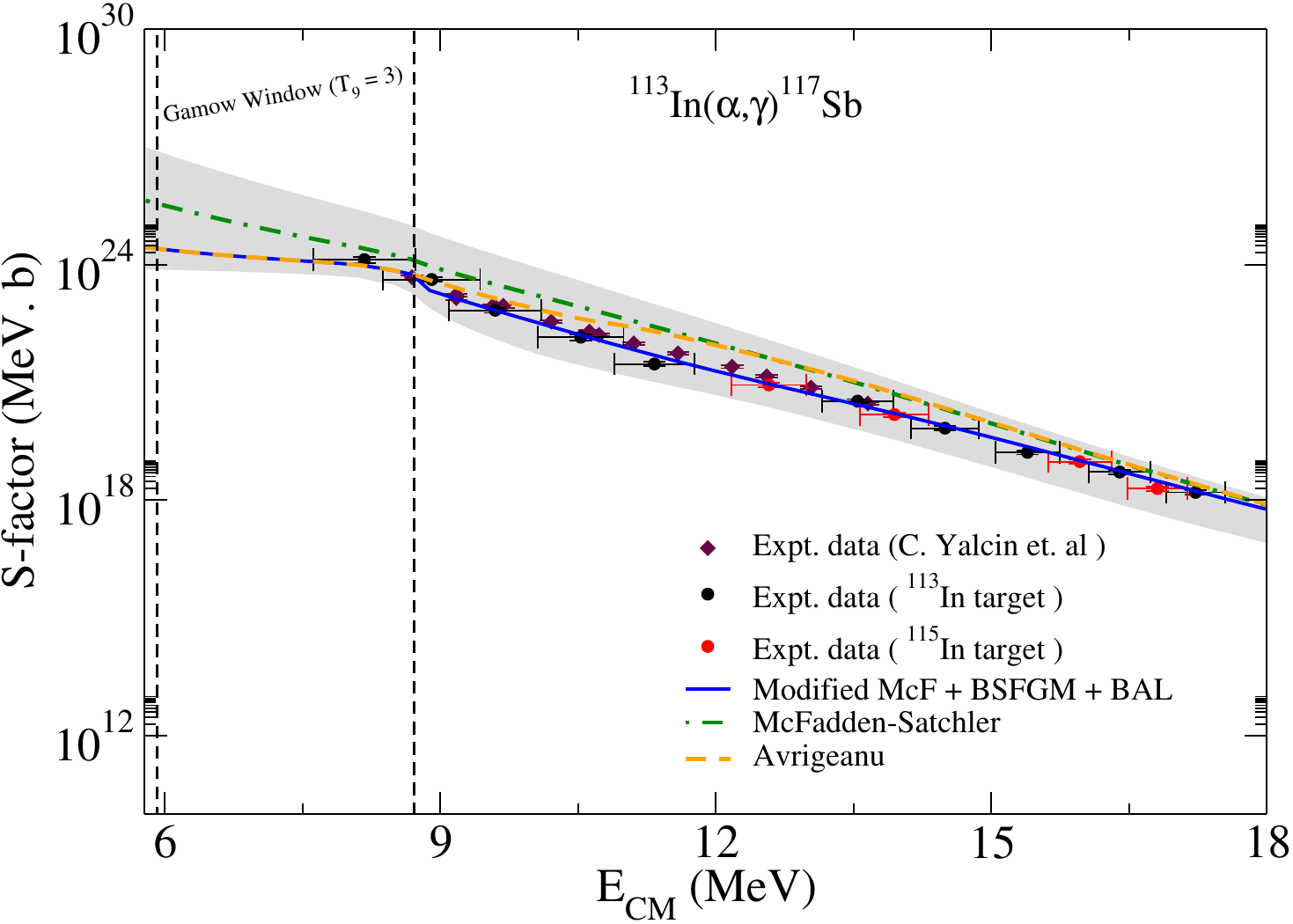}%
 \caption{\label{fig11}%
 Measured astrophysical $S$-factor of $^{113}$In($\alpha,\gamma$) reaction. Theoretical Hauser-Feshbach calculations have been performed using different nuclear input parameters. The gray shaded area imply that the theoretical calculation uncertainty using different nuclear input parameters ($\alpha$-OMP, LD, $\gamma$-SF). 
 }%
\end{figure}

\begin{figure*}[]
\includegraphics[scale=0.35]{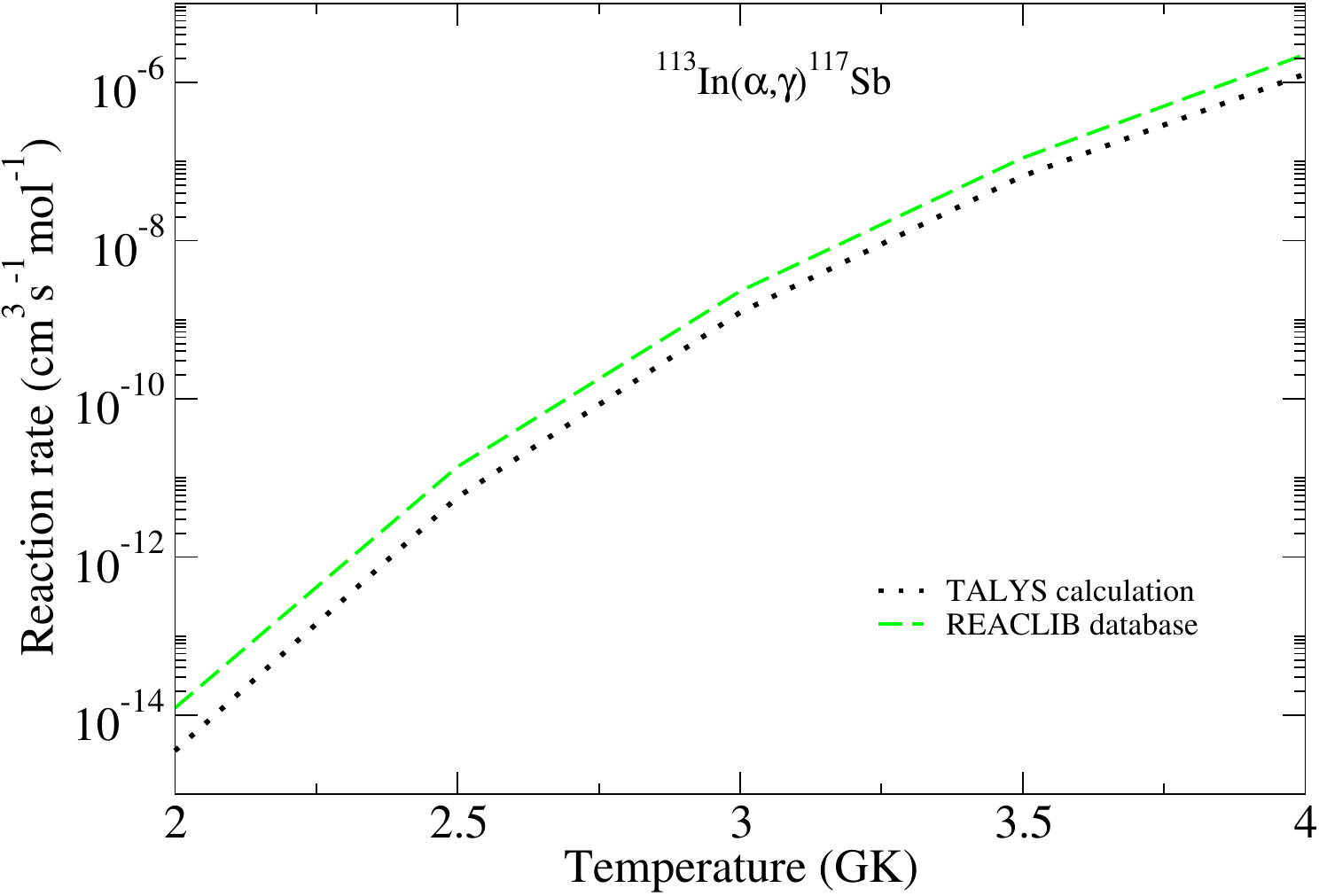}%
\includegraphics[scale=0.35]{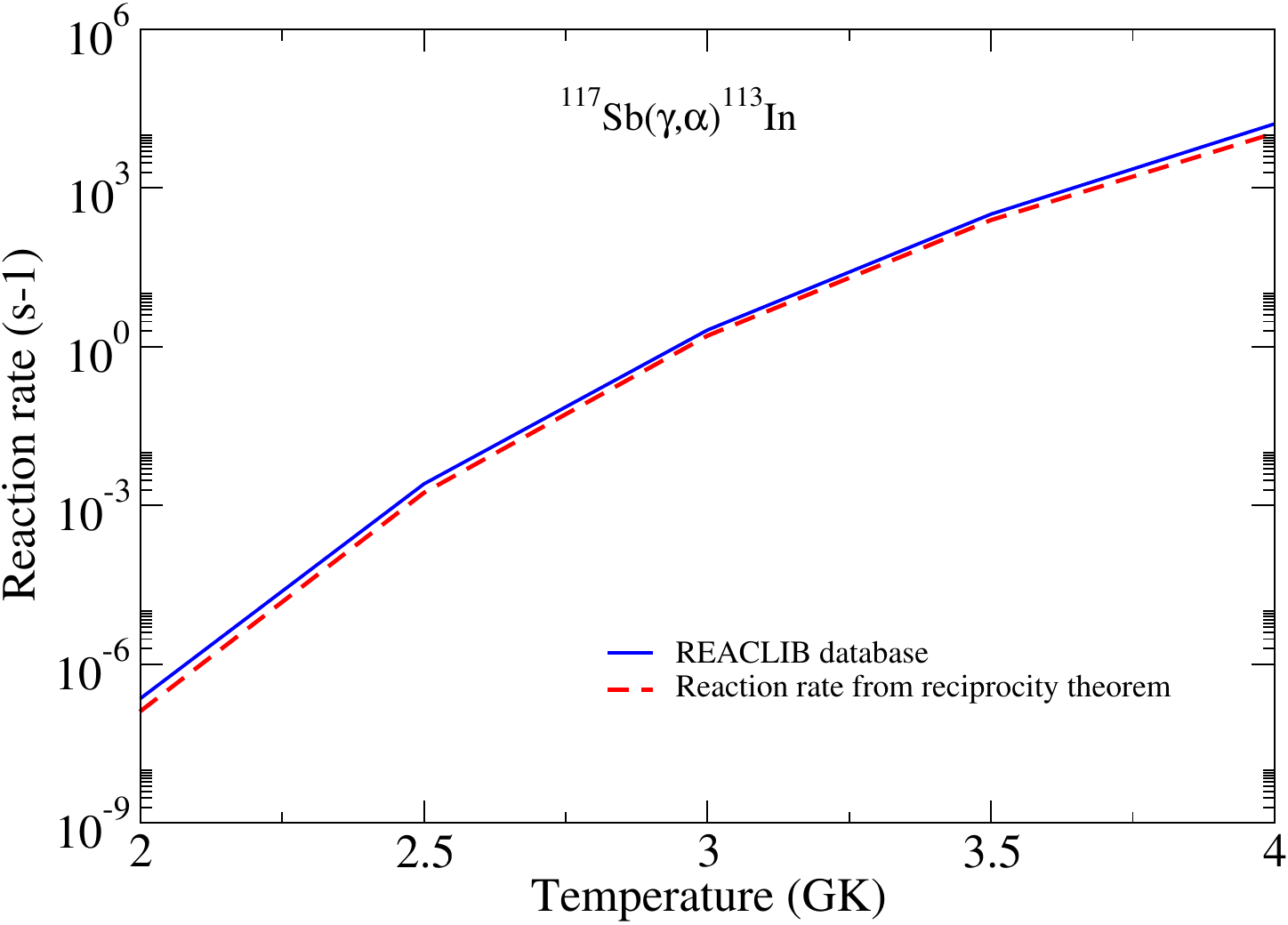}%
 \caption{\label{fig12}%
  $^{113}$In($\alpha,\gamma$)$^{117}$Sb and $^{117}$Sb($\gamma,\alpha$)$^{113}$In reaction rate at T$_9$ = 2$-$4 using the best-fitted nuclear input parameters (Modified McF, BFM, BAL) and compared with the reaction rates from REACLIB database~\cite{cyburt2010jina}.}%
\end{figure*}

The astrophysical $S$-factor is calculated by the given formula 
\begin{equation}
S({\rm E}) = {\rm E}\sigma({\rm E})e^{2\pi\eta}
\end{equation}
where E is center of mass energy, $\sigma$(E) is cross-section and  $\eta~=~\frac{Z_1Z_2e^2}{\hbar v}$ is the Sommerfeld parameter. $v$ being the relative velocity in the reaction and $\hbar$ is the reduced Plank constant. The measured $S$-factor was compared with the theoretical calculation, as shown in the Fig.~\ref{fig11}.
The ($\alpha,\gamma$) reaction rate in T$_9$ = 2$-$4 is calculated from the TALYS code using best-fitted nuclear input parameters. The stellar photodissociation rate of $^{117}$Sb($\gamma,\alpha$)$^{113}$In reaction is determined from the capture reaction rate $\left(N_{av}\bigl \langle \sigma v\bigr \rangle_{(\alpha,\gamma)}\right)$ using the reciprocity theorem~\cite{arnould2003p, holmes1976tables}
\begin{eqnarray}
\lambda_{(\gamma,\alpha)}(T) &=& \frac{(2J_{\rm{^{113}In}}^{0}+1)(2J_{\alpha}+1)}{(2J_{\rm{^{117}Sb}}^{0}+1)}~\frac{G_{\rm ^{113}In}^{0}}{G_{\rm ^{117}Sb}^{0}}~\left(\frac{A_{\rm{^{113}In}}A_{\alpha}}{A_{\rm{^{117}Sb}}}\right)^{3/2} \nonumber\\
& & \times
\left(\frac{kT}{2\pi\hbar^{2} N_{av}}\right)^{3/2}~ N_{av}\bigl \langle \sigma v\bigr \rangle_{(\alpha,\gamma)}~e^{-\frac{Q_{\alpha\gamma}}{kT}}
\end{eqnarray}
where $Q_{\alpha\gamma}$ is the Q-value of the $^{113}$In($\alpha,\gamma$) capture reaction and $k$ is the Boltzmann constant. $ N_{av}$ is the Avogadro number. $G_i^{0}$(T) is the temperature-dependent normalized partition function and $J_i^{0}$ is the ground state spin of nuclei $i$ with atomic mass number $A_i$.  $^{117}$Sb($\gamma,\alpha$)$^{113}$In and $^{113}$In($\alpha,\gamma$)$^{117}$Sb reaction rates are compared with the REACLIB database~\cite{cyburt2010jina} and shown in Fig.~\ref{fig12}. 

\section{Conclusions}
The $^{113}$In($\alpha,\gamma$)$^{117}$Sb and $^{113}$In($\alpha$, n)$^{116}$Sb$^{m}$ reaction cross-sections were measured using the stacked foil activation method. The reaction cross-section of $^{113}$In($\alpha,\gamma$) was measured for the first time at $E_{lab}$~=~8.46$\pm$0.58 MeV~($S$-factor = 1.34$\pm$0.24$\times$10$^{24}$ MeV.b). The measured ($\alpha,\gamma$) and ($\alpha$, n) reaction cross-sections were compared with the theoretical predictions using statistical model code TALYS-1.96 by varying different input parameters. Additionally, the sensitivity of different input parameters on the calculation of reaction cross-section was investigated. It was observed that $^{113}$In($\alpha$, n) reaction cross-section is well described by the modified McF $\alpha$-optical potential~\cite{basak2022determination}. The modified McF $\alpha$-optical potential provides good agreement of experimental $^{113}$In($\alpha,\gamma$) reaction cross-section and $S$-factor with level density model BFM  and $\gamma$-ray strength function BAL. However, based on sensitivity studies, the impact of level density and $\gamma$-ray strength function on $^{113}$In($\alpha,\gamma$) reaction is insignificant at astrophysical energies. The best-fitted input parameters were used to calculate the ($\alpha,\gamma$) reaction rate at T$_9$ = 2$-$4. $^{117}$Sb($\gamma, \alpha$) reaction rate was calculated from ($\alpha,\gamma$) reaction data using TALYS and reciprocity theorem. The measured ($\alpha,\gamma$) and ($\gamma, \alpha$) reaction rates differ from the REACLIB values by a factor of 2$-$4. More measurements at Gamow energies with $p$-nuclei are desired for a better understanding of the input parameters at astrophysical energies and therefore increase the predictability of the theoretical model. 

\section*{Acknowledgement}
The authors are immensely thankful to Dr. Chandana Bhattacharya for her invaluable assistance in completing the experiment successfully. The authors would like to thank Mr. A. A. Mallick of the Analytical Chemistry Division, BARC, VECC, and the crew of the K130 Cyclotron at VECC, Kolkata, for their generous support. Authors also acknowledge FRENA target facility for preparation of targets and Mr. Sudipta Barman and other workshop members of Saha Institute of Nuclear Physics, Kolkata for their kind support. SS would acknowledge the Council of Scientific and Industrial Research (CSIR), Government of India, for funding assistance (File No 09/489(0119)/2019-EMR-I).
\bibliography{dipali}

\end{document}